\documentclass[oneside,onecolumn,superscriptaddress,showpacs]{revtex4}
\usepackage{epsf}
\usepackage[dvips]{hyperref}
\begin{document}
\bibliographystyle{plain}
\nocite{*}
\title{Investigation of the fundamental constants stability based on the
reactor Oklo burn-up analysis}
\author{M.S. Onegin}
\email{onegin@thd.pnpi.spb.ru} \affiliation{St. Petersburg Nuclear
Physics Institute, Gatchina, 188 300, St. Petersburg, Russia}
\date{October 29, 2010}
\begin{abstract}

The burn-up for SC56-1472 sample of the natural Oklo reactor zone
3 was calculated using the modern Monte Carlo codes. We
reconstructed the neutron spectrum in the core by means of the
isotope ratios: $^{147}$Sm/$^{148}$Sm and $^{176}$Lu/$^{175}$Lu.
These ratios unambiguously determine the spectrum index and core
temperature. The effective neutron absorption cross section of
$^{149}$Sm calculated using this spectrum was compared with
experimental one. The disagreement between these two values allows
to limit a possible shift of the low laying resonance of
$^{149}$Sm even more . Then, these limits were converted to the
limits for the change of the fine structure constant $\alpha$. We
found that for the rate of $\alpha$ change the inequality $|\delta
\dot{\alpha}/\alpha| \le 5\cdot 10^{-18}$ is fulfilled, which is
of the next higher order than our previous limit.
\end{abstract}



\maketitle

\section{Introduction}

The confirmation of the temporal variation of the fundamental
constants would be the first indication of  the universe expansion
influence on the micro physics. One of the possibilities is
dependence of the nuclear couplings on the value of the cosmon
field - a possible candidate for the dark energy
(see~\cite{Wetterich1988,Ratra} and~\cite{Copeland} for review) -
the fundamental scalar field of high amplitude but tiny mass. If
the equation of this field state $w=p/\rho$ differs from -1 then
the phenomenology of this field will be different from the
universe with the cosmological constant. Then, the temporal
dynamics of the state equation could be determined by
investigation of the variation of fundamental constants with the
increased accuracy~\cite{Nunes}. The investigation of the natural
reactor Oklo allow to bound the possible variation of the fine
structure constant $\alpha$ with record high
accuracy~\cite{Olive,Shuryak} and the deviation of w from -1 now
($z\approx 0.13$)~\cite{Wetterich2003}.

Shlyakhter was the first who showed that the variation of the
fundamental constants could lead to measurable consequences on the
Sm isotops concentrations in the ancient reactor
waste~\cite{Shlyakhter}. Later Damur and
Dyson~\cite{Damour:1996zw} for Zones 2 and 5 and also
Fujii~\cite{Fujii:1998kn} for Zone 10 of reactor Oklo made more
realistic analysis of the possible shift of fundamental constants
during the last $~2\cdot 10^9$ years based on the isotope
concentrations in the rock samples of Oklo core. In this
investigation the idealized Maxwell spectrum of neutrons in the
core was used. The efforts to take into account more realistic
spectrum of neutrons in the core were  made in
works~\cite{Lamoreaux:2003ii,Petrov:2005pu,Lamur06}.

This paper continues the neutronics analysis of the Oklo reactor
that we began in the previous work~\cite{Petrov:2005pu}. In
paper~\cite{Petrov:2005pu} we considered a fresh core only and did
not analyze the method of extraction of $^{149}$Sm absorption
cross section by the geochemical analysis. Meanwhile, this cross
section depends on the reactor burn-up and on the neutron spectrum
transformation during the reactor operation. The bounds of the
$^{149}$Sm resonance shift during the past $~2\cdot 10^9$ years
have been obtained here in a different way as compared
to~\cite{Petrov:2005pu}. We used a modern reactor code to
investigate the burn-up of a particular sample from Zone 3 of
reactor Oklo. The conditions of the reactor operation could be
determined according to the distribution of different nuclide in
the sample. This procedure definitely determined the reactor
spectrum based on which the theoretical effective neutron
absorption cross section of the $^{149}$Sm nuclei can be
calculated. The obtained effective cross section have considerably
narrow uncertainties as compared to~\cite{Petrov:2005pu} where we
vary the uranium content, water fraction and temperature of the
reactor in the wide range. On the other hand, this cross section
can also be evaluated by the isotopic ratios of the Samarium
nuclide in the sample. Performing the fuel realistic burn-up we
can determine the cross section of $^{149}$Sm needed to describe
the concentration of this isotope in the sample. Comparing this
value with the theoretical one we can make a conclusion about the
variation of the $^{149}$Sm absorption cross section during the
time after the Oklo phenomenon occurred.

A similar approach was used in~\cite{Lamur06}. However, in this
paper the close limits on the temperature of the reactor core were
accepted. To our opinion, this point needs the further
investigation. Contrary, in our analysis  the temperature of the
reactor core during its operation is determined based on the
lutetium isotope ratio. Also, the water content in~\cite{Lamur06}
was not controlled. The uncertainties in the water concentration
in the sample also considerably influence the result.

\section{Determination of $^{149}$Sm effective neutron absorption cross
section}

We use the following definition of the effective cross
section~\cite{Damour:1996zw}
\begin{equation}
\label{sigSm}
\hat \sigma =\frac{\int \sigma (E) \Phi (E) dE}{v_0
\int n(E) dE},
\end{equation}
where $\Phi(E)$ - flux of neutrons with energy E, and
$n(E)=\Phi(E)/v$ - neutron density at energy $E$, $\sigma(E)$ -
the $^{149}$Sm nuclei absorption cross section of neutron. A
neutron with energy 0.0253 eV has velocity $v_0=2200$ m/sec. The
integral in~(\ref{sigSm}) is over the entire interval of neutron
energies. Equation~(\ref{sigSm}) determine the local instant
definition of the effective cross section because the neutron flux
$\Phi(E)$ is different for  the reactor core different areas and
for different periods of the reactor operation. As was claimed
earlier, this cross section strongly depends  on the core
temperature~\cite{Damour:1996zw,Fujii:1998kn,Petrov:2005pu,Lamoreaux:2003ii}
 and its element
composition~\cite{Lamoreaux:2003ii,Petrov:2005pu}.

\begin{figure}
\centerline{ \epsfxsize=15cm\epsfbox{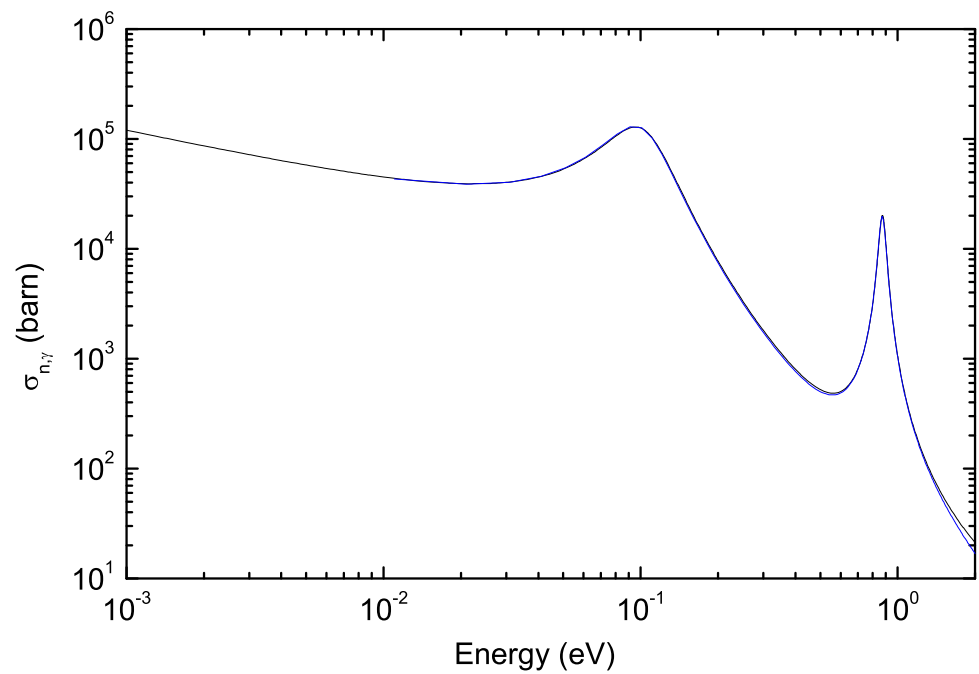}} \caption{ENDF
capture cross section of $^{149}$Sm - black solid line. Two
resonance approximation of capture cross section - blue solid
line.} \label{fig:GSm149}
\end{figure}

Neutron absorption cross section of $^{149}$Sm at low neutron
energies is saturated by the capture cross section. We use two
resonance approximations of the neutron capture cross section for
$^{149}$Sm by taken into account only the resonances at 97.3 meV
and 0.87 eV energies. The Breit-Wigner formula of the capture
cross section for one resonance as follows:

\begin{equation}
\label{BrWig} \sigma_{(n,\gamma)}(E)=g_0\frac{\pi \hbar ^2}{2 m E}\frac{\Gamma_n(E)
\cdot \Gamma_\gamma }{(E-E_\gamma)^2+\Gamma^2_{tot}/4}
\end{equation}
where $g_0$ is the statistical factor, $\Gamma_n, \Gamma_\gamma,
\Gamma_{tot}$ - neutron widths (see~\cite{Petrov:2005pu} for
resonance parameters). The calculated cross section is compared
with the evaluated absorption cross section for $^{149}$Sm taken
from Evaluated Nuclear Data File -ENDF-B/VI-8 (~\cite{ENDF}) in
Figure~\ref{fig:GSm149}. We can see that the law energy part of
cross section is described with~(\ref{BrWig}) rather well.

At first we investigated the influence of the uranium content in
the core, the amount of water in it and the core temperature on
the effective absorption cross section. We took the core
composition in accordance with our previous work
~\cite{Petrov:2005pu}. The Uranium weight percent in the dry ore
at the beginning of reactor operation was taken to be 38.4, 49.4
and 59.6\%. This values covered the uranium samples composition of
the Zone 2 that we investigated in this work. For example, the
uranium content of the samples for borehole SC36 of Zone 2(as
described by Ruffenach~\cite{Ruffenach}) are presented in
Table~\ref{Table1}.

\begin{table}[hb]
\caption{\label{Table1}}

\begin{tabular}{|c|c|c|}
\hline
 Sample & U content,\% & $^{235}$U/$^{238}$U \\
\hline
 SC36-1408/4 & 23.3 & 0.00545 \\
\hline
 SC36-1410/3 & 57.8 & 0.00527 \\
\hline
 SC36-1413/3 & 37.2 & 0.0041 \\
\hline
 SC36-1418 & 57.9 & 0.00574 \\
\hline
\end{tabular}
\end{table}

The core water content  in our calculations changed from 0.355
volume\% to 0.455 volume\%. We took the pressure of the core
during its operation to be 100 MPa. The composition of other ore
elements was given in~\cite{Petrov:2005pu}. This composition is
given in Table~\ref{Table2} for the dry ore in comparison with the
composition given in~\cite{Lamur06}. The difference between two
compositions is rather small and does not influence seriously the
calculated effective cross section.

\begin{table}[hb]
\caption{\label{Table2}Present day composition of the empty rock}
\begin{tabular}{|c|c|c|}
\hline
 Chemical composition & \% by weight (this work) & \% by weight (from~\cite{Lamur06})
 \\
 \hline
 SiO$_2$ & 43.00 & 42.14 \\
 \hline
 Al$_2$O$_3$ & 25.73 & 27.93 \\
 \hline
 FeO, Fe$_2$O$_3$ & 19.0 & 17.49 \\
 \hline
 MgO & 10.43 & 9.04 \\
 \hline
 MnO & - & 1.06 \\
 \hline
 K$_2$O & 1.84 & 2.35 \\
 \hline
\end{tabular}
\end{table}

\begin{figure}
\centerline{ \epsfxsize=15cm\epsfbox{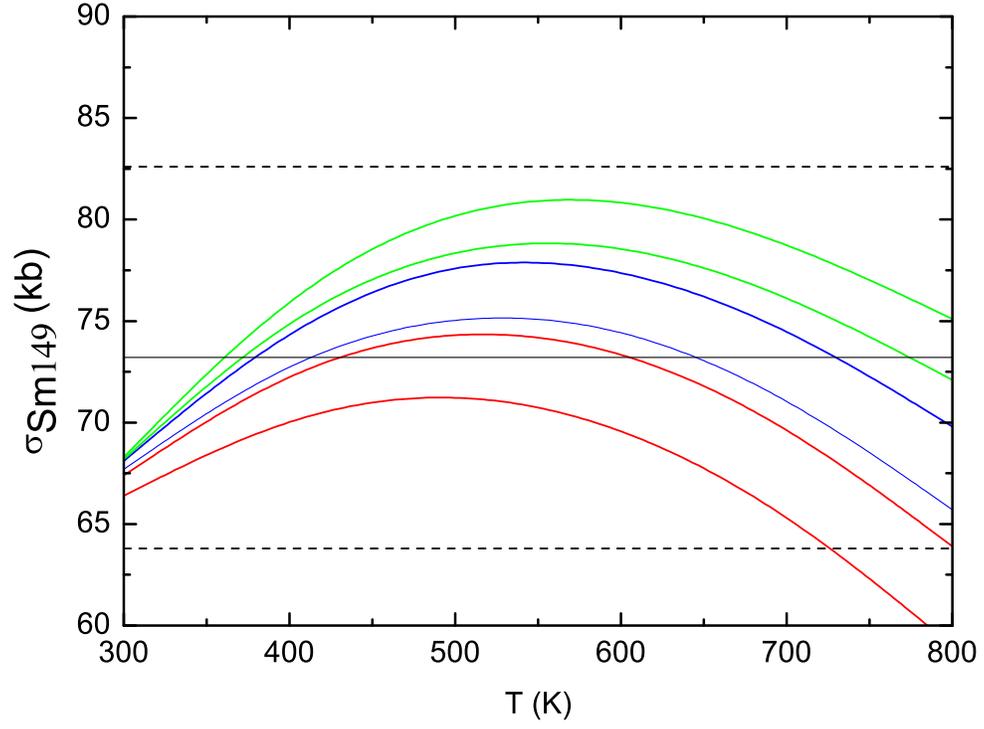}}
\caption{Dependence of effective capture cross section of
$^{149}$Sm on the core temperature. Green lines - uranium content
35\%, upper curve - $\omega_{H_2O}$=0.355, lower curve -
$\omega_{H_2O}$=0.455. Blue lines - uranium content 45\%, upper
curve - $\omega_{H_2O}$=0.355, lower curve -
$\omega_{H_2O}$=0.455. Red lines - uranium content 55\%, upper
curve - $\omega_{H_2O}$=0.355, lower curve -
$\omega_{H_2O}$=0.455. Also, the error corridor of
$\sigma_{Sm149}$ measured values from
work~\cite{Damour:1996zw}:$73.2\pm 9.4$ kb is shown. }
\label{GSm149Temp}
\end{figure}

We present in Figure~\ref{GSm149Temp} the effective cross section
of $^{149}$Sm calculated by  code MCNP-4C~\cite{MCNP} for
different temperatures of the fresh core and for different content
of uranium and water. The experimental (1 $\sigma$) bounds on the
measured cross section in different samples obtained in the
work~\cite{Damour:1996zw} are shown in the same Figure. The
calculation was performed with the Monte Carlo method for the
neutron spectrum which is set in the small cell.  Locally, the
neutron spectrum in the reactor can be approximated in the same
manner. Also, these results are presented in Table~\ref{Table3}.

\begin{table}[hb]
\caption{\label{Table3} Effective neutron cross section of
$^{149}$Sm}
\begin{tabular}{|c|c|c|}
 \hline
 & \multicolumn{2}{|c|}{$\hat \sigma_{149Sm}$} \\
 \cline{2-3}
 & \multicolumn{2}{|c|}{38.4 wt.\% U in ore}  \\
 \hline
 &$\omega_{H_2O}=0.355$
 &$\omega_{H_2O}=0.455$  \\
 300 & 68.2 & 68.3 \\
 400 & 75.5 & 76.6 \\
 500 & 78.9 & 80.8 \\
 600 & 79.0 & 81.3 \\
 700 & 76.4 & 79.0 \\
 800 & 72.1 & 75.1 \\
 \hline
 & \multicolumn{2}{|c|}{49.4 wt.\% U in ore}  \\
 \hline
 &$\omega_{H_2O}=0.355$ &$\omega_{H_2O}=0.455$  \\
 300 & 67.7 & 68.1 \\
 400 & 73.3 & 74.9 \\
 500 & 75.5 & 78.2 \\
 600 & 74.9 & 77.8 \\
 700 & 71.4 & 74.8 \\
 800 & 65.7 & 69.8 \\
 \hline
 & \multicolumn{2}{|c|}{59.6 wt.\% U in ore}  \\
 \hline
 & $\omega_{H_2O}=0.355$ &$\omega_{H_2O}=0.455$  \\
 300 & 66.4 & 67.4 \\
 400 & 70.5 & 72.8 \\
 500 & 71.7 & 74.8 \\
 600 & 70.0 & 73.8 \\
 700 & 65.7 & 70.0 \\
 800 & 59.0 & 63.9 \\
 \hline
\end{tabular}
\end{table}

The effective cross section is lowered with the increase of
uranium content in the core and with decreasing of the water
content. The temperature dependence of the cross section has the
maximum. (In the calculation the dependence of the water density
on temperature~\cite{Petrov:2005pu} was taken into account). The
position of the maximum is changed with the increasing of the
uranium amount. Experimental boundaries of Damour and Dyson with 2
$\sigma$ errors include all the calculated values. It shows that
the hypothesis about the constancy of effective absorption cross
section of $^{149}$Sm during the past 2 billion years is valid
with the equal precision. Nevertheless, this analysis was done for
the fresh core and the influence of fuel burn-up on the value of
the effective cross section needs further investigation. This will
be analyzed in the next section.

\section{Dependence of the effective cross section on burn-up}

\begin{figure}
\centerline{ \epsfxsize=15cm\epsfbox{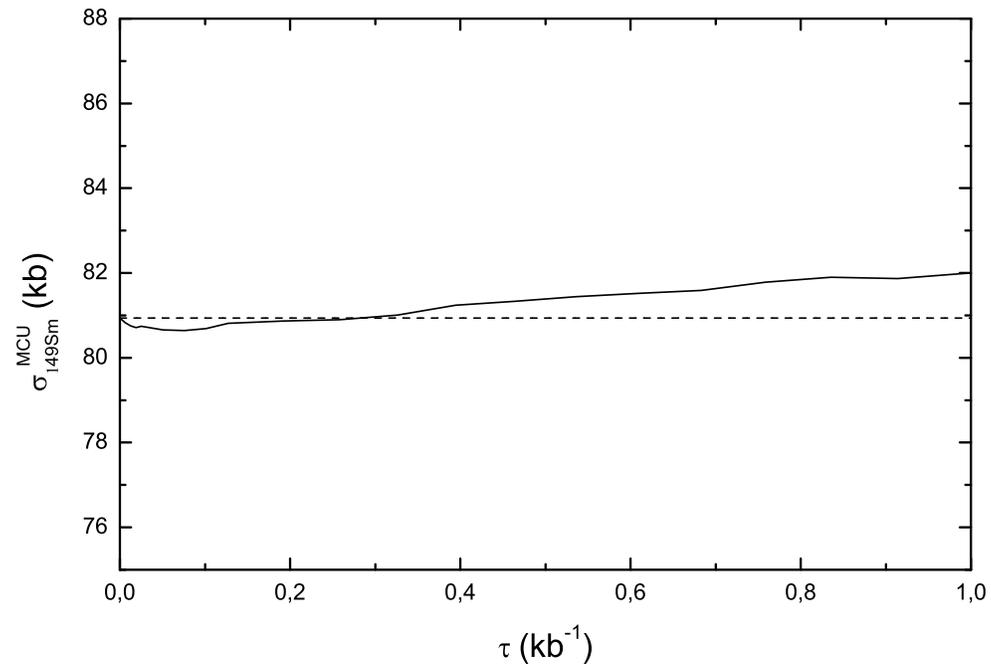}} \caption{
Dependence of the calculated effective capture cross section of
$^{149}$Sm on $\tau$.} \label{GSm149bur}
\end{figure}

During the reactor operation, the fuel products are accumulated in
the reactor core. Some products intensively absorb neutrons due to
its high absorption cross section. This causes the modification of
the reactor spectrum as far as the reactor burnt-up. To
investigate this issue, we performed  the realistic calculations
of burn-up with the help of  reactor code MCU-REA~\cite{MCU}. The
parameters of the reactor was taken to be closed to one in Zone 2:
the duration of reactor operation was about 800 thousand years,
the accepted reactor power - 25 kW ($1.22\cdot 10^{-4}$
W/cm$^3$)~\cite{Petrov:2005pu}. Kinetics of the concentrations of
actinides and fuel products were calculated according to the
following system of equations

\begin{equation}
\frac{dA_n}{dt}=-(\lambda_n+\hat \sigma_n \hat \phi)A_n+\sum_{i}\lambda_{i,n}A_i.
\end{equation}
Where $A_n$ - concentration on nuclide number $n$, $\hat \phi$ -
effective neutron flux, $\hat \sigma_n$ - effective absorption
cross section of $n$-th nuclide, $\lambda_n$ - its decay constant,
$\lambda_{i,n}$ - decay constant of nuclide number $i$ decay of
which lead to nuclide number $n$. Physical values of $\hat
\sigma_n$ and $\hat \phi$ were calculated in the nodes of the time
grid by the Monte Carlo method. In the calculation all complicated
physics of nuclear reactor theory was taken into account. That is
the resonance structure of cross sections of fuel and constituent
nuclides, the real reactor neutron spectrum established at the
time $t$ of the  reactor operation, the accumulation of fission
products and transuranium elements etc. The calculated dependence
of the effective cross section $\hat \sigma_{Sm149}$ on neutron
fluence $\tau$
$$\tau=\int \hat \phi(t) dt$$
is shown at Figure~\ref{GSm149bur}. In this calculation the
uranium content in the core was 49.4\% at the beginning of the
reactor operation, and the amount of water was
$\omega_{H_2O}=0.455$. The temperature of the core was taken to be
500K. At the beginning of the reactor operation the effective
cross section of $^{149}Sm$ was 80.94 kb and at the end of
operation ($\tau \approx 1.0$ kb$^{-1}$) it changed to 81.99 kb.
The change is about 1.3\%. For different core content and
different temperatures the situation is nearly the same - the
effective cross section of $^{149}Sm$ is nearly constant during
the reactor operation.  Thus, the core composition and the
temperature of the core  has the main effect on the value of the
effective absorption cross section of the Samarium. The spectrum
modification uncertainties due to fuel burn-up have a minor
importance.

\section{Sample SC52-1472}

Earlier it was shown that the main uncertainty in  determination
of the $^{149}$Sm effective absorption cross section contributes
the sample variation of the uranium amount, water content and the
core temperature. The main idea of the following investigation is
to restrict considerably and locally the conditions of the chain
reaction occurring. For this purpose we considered the burn-up of
sample SÑ52-1472 from the reactor Zone 3~\cite{Holliger}. This
approach is similar to one adopted in~\cite{Lamur06}. In this
paper, the burn-up of the samples from the Zone 2 and Zone 10 were
analyzed. The main difference of our approach from one
in~\cite{Lamur06} is that we fix the amount of water in the sample
during the reaction operation and the temperature of the neutron
spectrum. Also, the Samarium amount in the sample was determined
before the beginning of the chain reaction.

For the particular sample we definitely know the content of the
uranium in it and initial composition of the rear earth
impurities. We know the burn-up of the uranium in the sample at
the end of the reactor operation. The duration of the chain
reaction can be determined if the contribution of the $^{238}$U
($\alpha$) and $^{239}$Pu ($\beta$) nucleus to fission is known.
This contribution can be determined from the following nuclear
isotope ratios in the sample:

\begin{tabular}{c}
$^{150}$Nd/($^{143}$Nd+$^{144}$Nd+$^{145}$Nd+$^{146}$Nd), \\
$^{154}$Sm/($^{147}$Sm+$^{148}$Sm), \\
($^{157}$Gd+$^{158}$Gd)/($^{155}$Gd+$^{156}$Gd).\\
\end{tabular}

If follows from the fact that the yield of isotopes $^{150}$Nd,
$^{154}$Sm and $^{157}$Gd,$^{158}$Gd in fission strongly depends
on the nuclei that undergoes fission. The calculation shows that
for a particular sample about 2\% of fissions were due to
$^{238}$U and $^{239}$Pu.

The production of the $^{239}$Pu nuclei in the reaction of neutron
absorption by the $^{238}$U nuclei also influences the reactor
energy balance.  The Plutonium isotope have the half life equal to
24110 years and transmitted into the $^{235}$U isotope after the
$\alpha$-decay. We identify the conversion factor as
$C$~\cite{Damour:1996zw}. The duration of reactor operation can be
evaluated then as follows:

\begin{equation}
\label{duration} \Delta t= \frac{\hat \sigma_{239,f}\hat
\sigma_{235}}{\hat \sigma_{235,f}}
\frac{1-\alpha-\beta}{\beta}\frac{C\tau}{\lambda_{239}}.
\end{equation}
In this equation $\hat \sigma_{235,f},\hat \sigma_{239,f}$ -- is
the effective fission cross section of $^{235}$U and $^{239}$Pu
respectively, $\hat \sigma_{235}$ -- absorption cross section of
$^{235}$U, $\lambda_{239}$ - decay constant of $^{239}$Pu nuclei.
Based on the produced energy in a volume unit and the duration of
the nuclear chain reaction the mean nuclear power release in the
sample can be evaluated.

The main factor that influences the spectrum index is the amount
of water in the sample~\cite{Lamur06}. The very precise measure of
the spectrum index is the value of the $^{147}$Sm  absorption
cross section. The approximate relation between the effective
absorption cross section of the $^{147}$Sm and the spectrum index
$r$  as follows:
$$\hat \sigma _{Sm147}=0.88\, r,$$
where cross section is in kilo barns. The absorption cross section
of the $^{147}$Sm isotope, on the other hand,  influences  the
isotopic ratios of Sm in the sample, notably the content of
$^{148}$Sm, because this isotope is not produced within the fuel
fission. Varying the amount of water in the sample we can
reproduce the experimental ratio for the $^{148}$Sm isotope. Thus,
the procedure fixes the spectrum index.

\begin{figure}
\centerline{ \epsfxsize=15cm\epsfbox{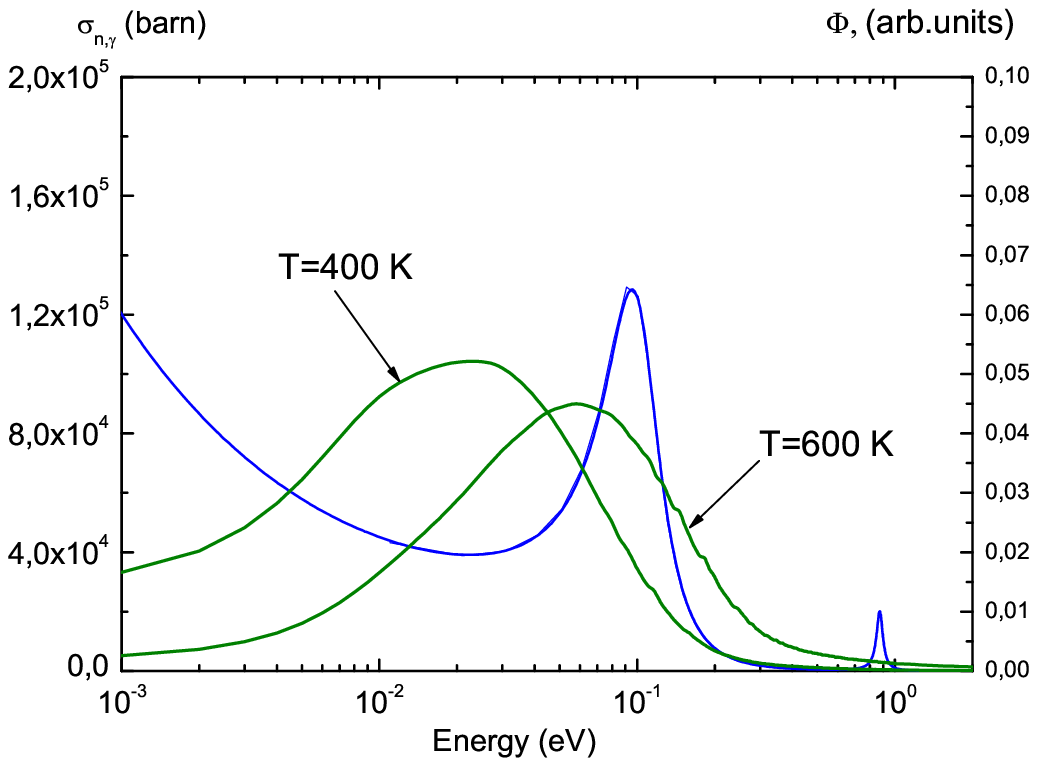}} \caption{Energy
dependence of the effective absorption cross section of the
$^{149}$Sm (left, blue line). Calculated neutron fluxes for
temperatures 400 and 600 K (right, green lines).} \label{Sig149T}
\end{figure}

\begin{figure}
\centerline{ \epsfxsize=15cm\epsfbox{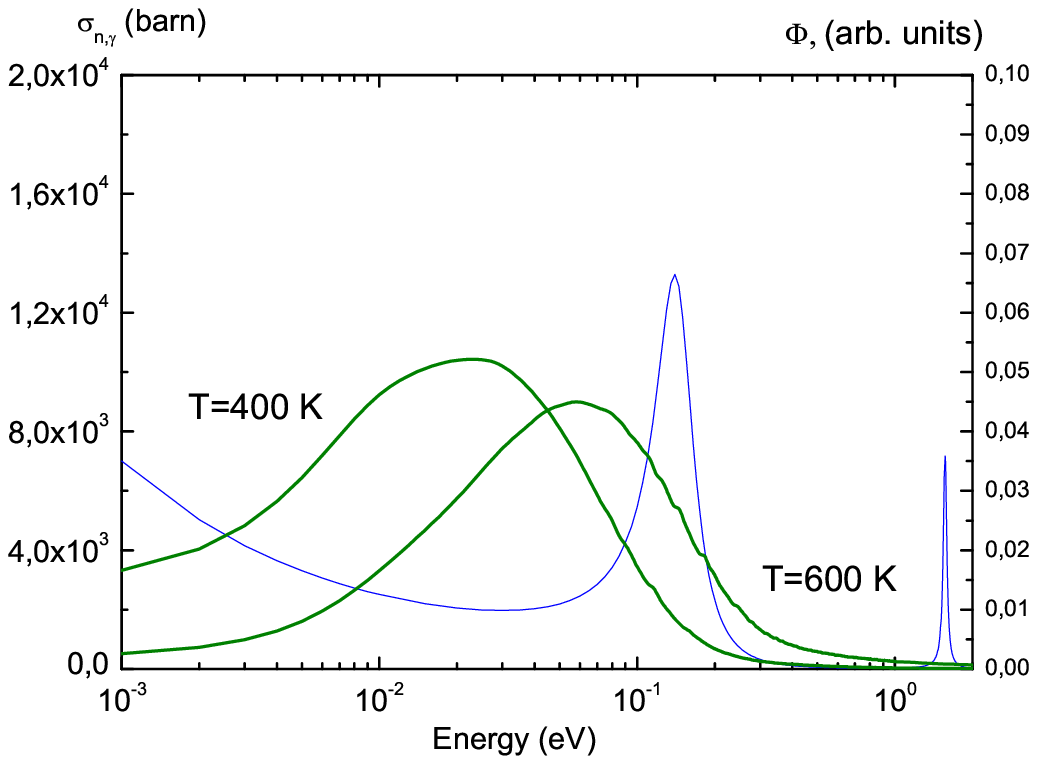}}\caption{Energy
dependence of the effective absorption cross section of the
$^{176}$Lu (left, blue line). Calculated neutron fluxes for
temperatures 400 and 600 K (right, green lines).} \label{Lu176abs}
\end{figure}

The next unknown quantity is the core temperature. The neutron
spectrum in reactor significantly depends on the core temperature.
The higher temperatures, the more probable higher neutron
energies. In Figure~\ref{Sig149T} we present two neutron spectrums
calculated according to code MCNP-4C for T = 400 and 600 Ê as
compared to the effective absorption cross section by $^{149}$Sm.
To fix the temperature we used the experimental ratio of two Lu
isotopes in the sample: $^{176}$Lu/$^{175}$Lu. As for a $^{149}$Sm
isotope, the $^{176}$Lu isotope has  a low energy resonance. The
energy of the resonance is 141.3 meV. The absorption cross section
is plotted in Figure~\ref{Lu176abs} in comparison with two neutron
spectrums with Ò = 400 and 600~K. As you can see from the
comparison, the neutron spectrum shifted to the right, when its
temperature increased considerably overlapping the resonance cross
section for higher temperature. Determining the spectrum
temperature from the description of the Lu isotope ratio in the
sample and the spectrum index from the description of the
$^{148}$Sm isotope ratio we mainly determined the form of the
neutron spectrum in the equation (\ref{sigSm}). Using the
$^{149}$Sm cross section from (\ref{BrWig}) we can determine the
{\it theoretical effective cross section} of this isotope.

The main parameters of the sample and the time parameters of the
reactor in Zone 3 are presented in
Table~\ref{Table4}~\cite{Holliger,Holliger2}. The nuclide
composition of the sample (adopted in our calculations) at the
beginning of the burn-up are presented in Table~\ref{Table5}. The
density of the dry ore was determined from content of the Uranium
in it (see~\cite{Petrov:2005pu}). It was equal to 3.465 g/cm$^3$.
The isotope ratio of Uranium at the beginning of the reactor
operation was determined using the age of the reactor and the
current normal isotope ratio for Uranium. At the beginning of the
reactor life this ratio was equal to: $^{235}$U/$^{238}$U = 3.6\%.
The composition of the empty rock was calculated in accordance
with the Table 2. The amount of Samarium in the sample before
operation was determined based on its current concentration using
the fact that $^{144}$Sm isotope is not produced in fission but
occurred in the sample before the beginning of the chain reaction.
The other isotopes were distributed according to normal isotopic
ratio. The same procedure was applied to Neodymium but based on
the concentration of the $^{142}$Nd isotope. We neglected the
chemical reshuffling after the chain reaction ended.

\begin{table}[hb]
\caption{\label{Table4}}

\begin{tabular}{|c|c|c|c|c|}
 \hline
 U content  & Fluence,  & Duration,  & Age, MY & Specific power,  \\
in sample, wt.\% & kb$^{-1}$ & years & & W/cm$^3$ \\
 \hline
 $0.32 \pm 0.01$ & 0.228 & 300 000 & 1.93 & $0.45\cdot 10^{-4}$  \\
 \hline
 \end{tabular}
\end{table}

\begin{table}
\caption{\label{Table5}Initial density of SC56-1472 sample}

\begin{tabular}{|c|c|c|c|c|c|}
 \hline
Element  & Concentration, & Element  & Concentration, &Element  & Concentration, \\
 & atoms/barn cm & & atoms/barn cm & & atoms/barn cm \\
\hline
 U-235 & $0.10751 \cdot 10^{-3}$ & Nd-142 & $0.20149 \cdot 10^{-6}$ & Sm-144 & $3.186 \cdot 10^{-9}$ \\
 U-238 & $0.29898 \cdot 10^{-2}$ & Nd-143 & $0.90376 \cdot 10^{-7}$ & Sm-147 & $0.14839 \cdot 10^{-7}$ \\
 O     & 0.058581                & Nd-144 & $0.17631 \cdot 10^{-6}$ & Sm-148 & $0.11126 \cdot 10^{-7}$ \\
 H     & 0.046799                & Nd-145 & $0.61485 \cdot 10^{-7}$ & Sm-149 & $0.13680 \cdot 10^{-7}$ \\
 Si    & $0.75364 \cdot 10^{-2}$ & Nd-146 & $0.12742 \cdot 10^{-6}$ & Sm-150 & $0.73055 \cdot 10^{-8}$ \\
 Al    & $0.53157 \cdot 10^{-2}$ & Nd-148 & $0.42225 \cdot 10^{-7}$ & Sm-152 & $0.26480 \cdot 10^{-7}$ \\
 Mg    & $0.27296 \cdot 10^{-2}$ & Nd-150 & $0.41484 \cdot 10^{-7}$ & Sm-154 & $0.22520 \cdot 10^{-7}$ \\
 K     & $0.41208 \cdot 10^{-3}$ & & & &  \\
 Fe    & $0.27191 \cdot 10^{-2}$ & & & &  \\
 \hline
 \end{tabular}
\end{table}

The value of the {\it experimental cross section} for $^{149}$Sm
can be determined from the description of the $^{149}$Sm isotope
concentration in the sample. As the sample nuclide composition was
fixed, we could change only the temperature of the core to vary
the concentration of this isotope at the end of the burn-up. We
have made three computations of the sample burn-up at three
different temperatures: 300, 400 and 500 K. We used the burn-up
utility MURE~\cite{mure} which is based on MCNP code. The isotope
ratios of the Sm at the end of the burn-up is presented in
Table~\ref{Table6} in comparison with the experimental data.
Table~\ref{Table6} confirms that the temperature of the core
should be equal to 367 Ê (using linear fit) to describe the
experimental concentration of the $^{149}$Sm in the sample. In
Table~\ref{Table7} effective cross section of $^{149}$Sm for this
temperatures is presented. From this table it is followed that the
cross section of $^{149}$Sm at 367 K is equal to 62.91 kb. Thus,
the value of the experimental effective absorption cross section
we determined is equal to 62.91 kb. This value can be compared
with the value obtained for considered sample SÑ56-1472
in~\cite{Damour:1996zw}. In this work for this cross section the
value 72 kbarn was received. Thus, our experimental cross section
of $^{149}$Sm is considerably lower than in~\cite{Damour:1996zw}.
This disagreement can be attributed to the procedure used
in~\cite{Damour:1996zw} to obtain an experimental cross section.
In this analysis the contribution of $^{238}$U and $^{239}$Pu to
fission was not taken into account. The influence of this nuclei
could be rather strong, and, if we take them into account, the
experimental cross section would be considerably smaller.

In Table~\ref{Table8}, the calculated isotope ratios for Neodymium
nuclei in the sample at the end of burn-up are compared to the
experimental data. The agreement is rather good. Particularly, the
experimental isotope ratio for the $^{150}$Nd isotope is
reproduced in our calculations. As you can see from the table, the
temperature dependence of isotopic ratios for Neodymium is weak.

\begin{table}[hb]
\caption{\label{Table6}Calculated isotopic ratios of Sm at the end
of burn-up}

\begin{tabular}{|c|c|c|c|c|}
 \hline
Sm isotope & \multicolumn{3}{|c|}{Temperature, K} & Exp. distrib.\\
 \cline{2-4}
 & 300 & 400 & 500 & \\
 \hline
144 & 0.53 & 0.54 & 0.53 & $0.53 \pm 0.005$ \\
147 & 50.93 & 51.29 & 50.99 & $51.55 \pm 0.06$ \\
148 & 2.70 & 2.74 & 2.71 & $2.73 \pm 0.01$ \\
149 & 1.47 & 1.25 & 1.19 & $1.32 \pm 0.005$ \\
150 & 25.60 & 25.58 & 25.93 & $25.36 \pm 0.03$ \\
152 & 13.31 & 13.10 & 13.16 & $12.94 \pm 0.01$ \\
154 & 5.46 & 5.51 & 5.48 & $5.57 \pm 0.01$ \\
 \hline
\end{tabular}
\end{table}

\begin{table}[hb!]
\caption{\label{Table7}Effective cross section of $^{149}$Sm and
$^{175,176}$Lu for different temperatures}

\begin{tabular}{|c|c|c|c|}
 \hline
Isotope &\multicolumn{3}{|c|}{Temperature, K} \\
\cline{2-4}
 & 300 & 400 & 500  \\
 \hline
Sm149 & 55.67 & 66.47 & 72.49 \\
Lu175 & 0.0841 & 0.0869 & 0.0845 \\
Lu176 & 4.198 & 5.232 & 6.299 \\
 \hline
\end{tabular}
\end{table}

\begin{table}[hb]
\caption{\label{Table8} Calculated isotopic ratios of Nd at the
end of burn-up}

\begin{tabular}{|c|c|c|c|c|}
 \hline
Nd isotope & \multicolumn{3}{|c|}{Temperature, K} & Exp. distrib.\\
 \cline{2-4}
 & 300 & 400 & 500 & \\
 \hline
142 & 5.83 & 5.84 & 5.84 & $5.75 \pm 0.02$ \\
143 & 24.17 & 24.17 & 24.16 & $24.33 \pm 0.04$ \\
144 & 26.85 & 26.77 & 26.85 & $26.84 \pm 0.04$ \\
145 & 16.57 & 16.62 & 16.57 & $16.57 \pm 0.03$ \\
146 & 15.23 & 15.21 & 15.23 & $15.15 \pm 0.03$ \\
148 & 7.61 & 7.66 & 7.62 & $7.62 \pm 0.02$ \\
150 & 3.74 & 3.73 & 3.74 & $3.73 \pm 0.01$ \\
 \hline
\end{tabular}
\end{table}

Independently we can determine the temperature of the core during
the reactor operation from the lutetium isotope ratio in the
sample. Effective cross sections of lutetium isotopes at different
temperatures are presented in Table~\ref{Table7}. At the beginning
of the reactor operation we took the lutetium isotope ratio to be
normal:

$^{176}$Lu/$^{175}$Lu = 0.0266.

\begin{figure}
\centerline{ \epsfxsize=15cm\epsfbox{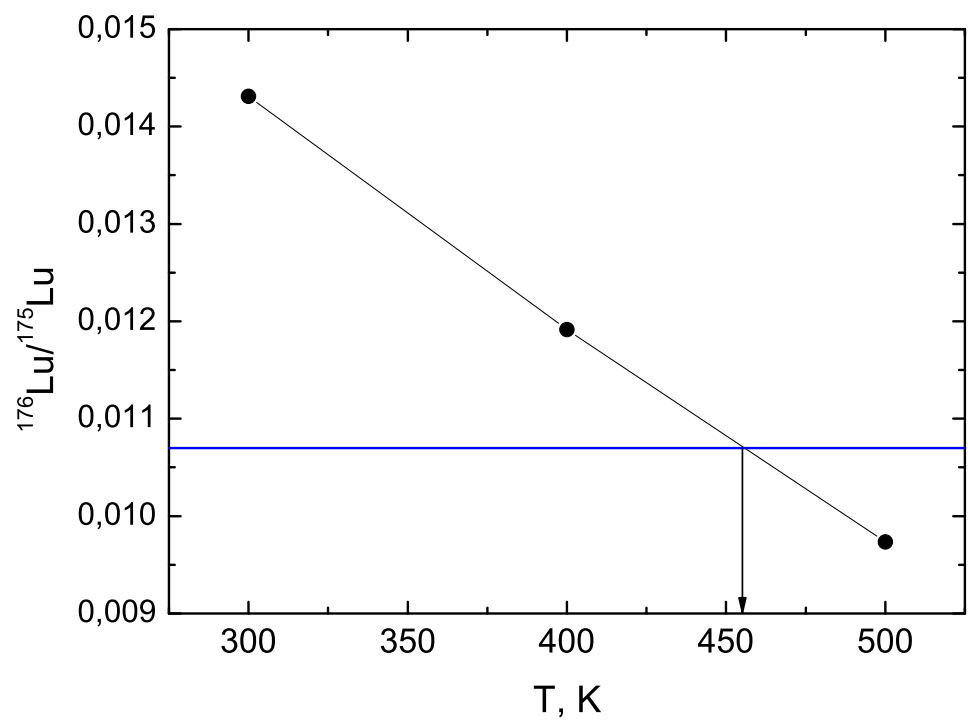}}
\caption{Dependence of $^{176}$Lu/$^{175}$Lu ratio at the end of
burn up cycle on the temperature.} \label{Luratio}
\end{figure}

The absorption of neutron by the $^{175}$Lu isotope leads to the
isotope $^{176}$Lu. Part of the $^{176}$Lu nuclei in this reaction
is produced in the metastable state. The spin of this metastable
state considerably differ from the spin of the ground state. Due
to this fact, the radiation transition from the metastable state
to ground one is forbidden and the main mode of decay of the
metastable state is the $\beta^-$  decay to $^{176}$Hf nuclei. The
total experimental absorption cross section of the thermal neutron
by $^{175}$Lu is equal to $23.3 \pm 1.1$ b~\cite{Muhabhab}. The
experimental cross section of production of $^{176}$Lu in the
ground state is equal to $6.6 \pm 1.3$ b~\cite{Muhabhab}. So the
experimental yield of the ground state in the neutron absorption
by $^{175}$Lu nuclei is equal to $0.283 \pm 0.056$ with $1 \sigma$
error. The yield of lutetium isotopes in fission is negligibly
small. The $^{175}$Lu nuclei can be produced only in the reaction
of neutron absorption with $^{174}$Yb nuclei. The effective
absorption cross section of neutron by $^{176}$Lu nuclei is
presented in Table~\ref{Table7}. It increases within the
temperature increasing. At higher temperatures $^{176}$Lu nuclear
burns up better in the sample. So the ratio $^{176}$Lu/$^{175}$Lu
in the sample can help to determine the spectrum temperature. The
dependence of this ratio at the end of burn-up cycle on the core
temperature  is presented in Figure~\ref{Luratio}. In good
approximation this dependence is linear. Experimentally this ratio
is equal to $^{176}$Lu/$^{175}$Lu$ = 0.0107 \pm
0.0001$~\cite{Holliger}. In Figure~\ref{Luratio} we can determine
what was the core temperature during the reactor operation. It's
equal to 455 K. The cross section of the $^{149}$Sm at this
temperature amounts to 69.78 kb from Table 6. The disagreement
between the $^{149}$Sm experimental effective cross section
obtained from the concentration of this isotope in the sample and
the theoretical value of this cross section calculated with the
help of the evaluated neutron spectrum is equal to 6.9 kb.

\begin{figure}
\centerline{ \epsfxsize=15cm\epsfbox{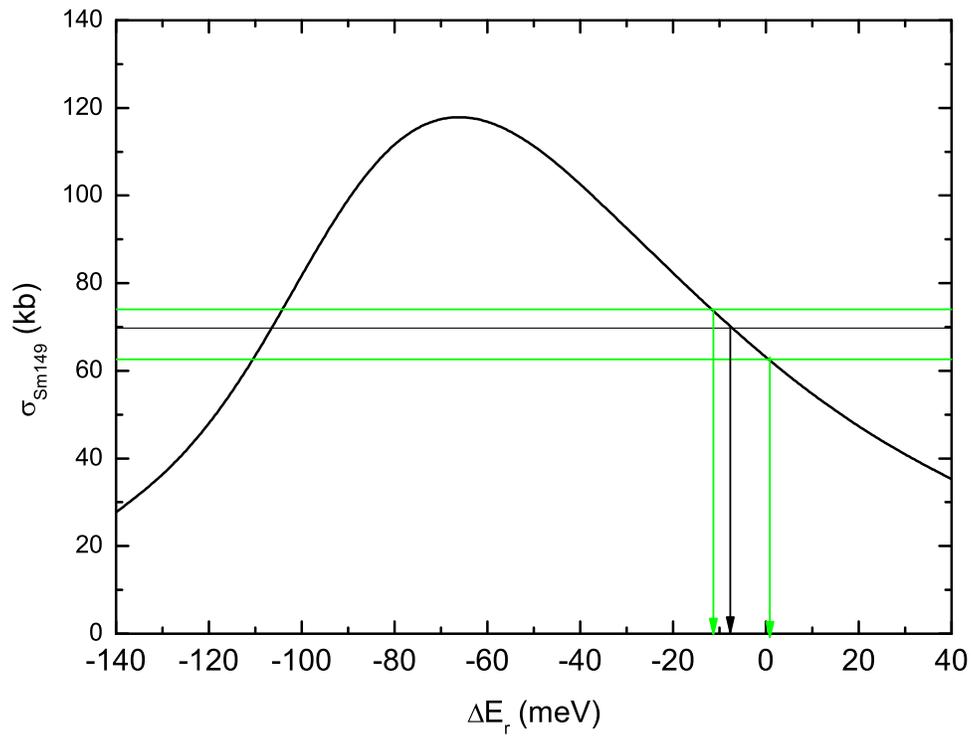}}
\caption{Dependence of the $^{149}$Sm cross section on the
resonance shift. } \label{Sg49vsEr}
\end{figure}

 Such disagreement in the $^{149}$Sm cross section experimental and
theoretical values can be explained with the help of the
hypotheses that the shift of the position of the $^{149}$Sm low
laying resonance occurred for about the past $2\cdot 10^9$ years
time. The dependence of the $^{149}$Sm cross section on the shift
of the resonance $\Delta E_r$ is shown in  Figure~\ref{Sg49vsEr}.
This dependence was calculated for the core temperature  equal to
400 K at the end of the burn-up for the real neutron spectrum. As
you can see in the Figure,  the theoretical value of the
$^{149}$Sm cross section equal to 69.78 kb can be explained by the
shift of the resonance position by amount $\Delta E_r=-7.6$ meV.
However, we must take into account uncertainties in the
temperature determined with the help of lutetium. This analysis is
performed in the next section.

\section{Analysis of the ambiguities}

The lutetium isotope ratio at the end of burn up is sensitive to
the ground state $^{176}$Lu yield in the absorption of neutron by
$^{175}$Lu nuclei. This yield have rather high uncertainty:~$0.283
\pm 0.056$ with $1 \sigma$  error. We have investigated the
dependence of the $^{176}$Lu/$^{175}$Lu ratio at the end of
burn-up on this yield. For the yield $0.283 - 3\cdot 0.056$ we
obtained the temperature equal to 364 K. On the other side, for
the yield $0.283 + 3\cdot 0.056$ the temperature is to be equal to
525 K. As a result the core temperature determined from the
lutetium isotope ratio is in the range
$$364 {\rm \,K} < T < 525 {\rm \,K}$$
at 99\% C.L. The values of the $^{149}$Sm cross section (as it
follows from Table 7) are in the range 62.6 kb to 74.0 kb.
Following the dependence of the $^{149}$Sm cross section on a
resonance shift (see Figure~\ref{Sg49vsEr}) we conclude that the
resonance shift should be  in the range
$$ -11.3 {\rm \,meV} < \Delta E_r < 0.8  {\rm \,meV},$$
to satisfy the experimental condition. So the zero resonance shift
is not excluded.

\section{Discussion of the results}

Constraints for the shift of the $^{149}$Sm resonance obtained in
the previous section can be converted into the limits for the
variation of the fine structure constant $\alpha$
(see~\cite{Petrov:2005pu})
\begin{equation}
\frac{\Delta \alpha}{\alpha}=\frac{\Delta E_r}{M}
\end{equation}
where: $M$ is estimated in~\cite{Damour:1996zw} as -$(1.1\pm 0.1)$
MeV. Using this value of $M$ we obtain constraints for the
variation of the fine structure constant
\begin{equation}
\label{varalpha} -0.7\cdot 10^{-9} \le \frac{\Delta
\alpha}{\alpha} \le 1.0\cdot 10^{-8}
\end{equation}
during the past $2\cdot 10^9$ years. Dividing the
equation~(\ref{varalpha}) by the age of the reactor $1.93\cdot
10^9$ years we get the following limit (99\% C.L.) for the time
derivative of $\alpha$
\begin{equation}
-0.35\cdot 10^{-18}{\rm \,\, yr}^{-1} \le \frac{\Delta \dot
\alpha} {\alpha} \le 5\cdot 10^{-18} {\rm \,\, yr}^{-1}.
\end{equation}
This constraint is of the next higher order than in our previous
work~\cite{Petrov:2005pu}.

The only evidences obtained till now about the variation of
fundamental constants include astrophysical observations of the
variation of the fine structure constant for $0.2 \le z \le 4.2$
and the change of the electron to proton mass ratio $\mu$ for
$z\sim 2.8 $~\cite{Murphy,Levshakov,Ivanchik}. If this variations
are due to fundamental scalar field, it causes also the violation
of the Week Equivalence Principle (WEP) of General
Relativity~\cite{Bekenstein,Olive2002,Will, Dent}. The reported
values of $\alpha$ and $\mu$ variation in astrophysical
observations, if they are true, predict the violation of WEP on
the observable level in future experiments.

Supersimmetric scenarios of the unifications of forces naturally
lead to the fundamental cosmon scalar field coupling with the
matter and electromagnetic energy. The variation of scalar field
over the cosmological times in the unified scenarios leads to the
variation of nuclear couplings. Consistency of different scenarios
of unification with the reported variations of the fundamental
constants and with the constraints for the violation of WEP was
analyzed in~\cite{Olive2002,Wetterich2003,Dent,Copeland2004}. In
this analysis the Oklo`s constrain for the variation of $\alpha$
serves as an additional rigid boundary condition to the evolution
of the cosmon field and generally provides the severely
restriction  for the possible unification couplings and
unification models.

In this paper we take into account only the influence of coulomb
energy on the position of resonances. It is highly unrealistic
that the modification of fine structure constant is accompanied
with changing of strong coupling constant is such a manner that
the total nuclear energy is unchanged. Contrary it can be expected
that the constrains will be ever harder~\cite{Olive}, if we admit
the unification forces hypotheses.

As a result, we state that the variation of $\alpha$ during the
past 2 billion years is not larger than $1.0\cdot 10^{-8}$. For
this result confirmation additional investigation of burn-up of
the reactor Oklo other samples is necessary

\section{Acknowledgments}

The author expresses his appreciation to E.A. Gomin, M.I.
Gurevich, A.S. Kalugin and M.S.~Yudkevich for providing the code
MCU-REA. The author feels it his pleasant duty to thanks M.S.
Yudkevich for discussions and consultation.


\begin{thebibliography}{99}

\bibitem{Wetterich1988}
C.~Wetterich, Nucl. Phys. {\bf B}302, 668 (1988)

\bibitem{Ratra}
B. Ratra and J. Peebles, Phys. Rev. {\bf D} 37, 321 (1988)

\bibitem{Copeland}
E.J. Copeland, M. Sami and S.~Tsujikawa, hep-th/0603057

\bibitem{Nunes}
P.P. Avelino, C.J.A.P~Martins,N.J.~Nunes and K.A.~Olive, Phys.
Rev. {\bf D} 74:083508,(2006); hep-ph/0605690

\bibitem{Olive}
K. A. Olive, M. Pospelov, Y.-Z. Qian, A. Coc, M. Casse, and E. Vangioni-Flam, Phys.
Rev. {\bf D66}, 045022 (2002); hep-ph/0205269

\bibitem{Shuryak}
V.V. Flambaum and E.V. Shuryak, Phys. Rev. {\bf D65}, 103503 (2002)

\bibitem{Wetterich2003}
C.~Wetterich, JCAP {\bf 0310}, 002 (2003) ; hep-ph/0203266

\bibitem{Shlyakhter}
A. Shlyakhter, Nature {\bf 264}, 340 (1976)

\bibitem{Damour:1996zw}
Thibault Damour and Freeman Dyson.
\newblock The oklo bound on the time variation of the fine-structure constant
  revisited.
\newblock {\em Nucl. Phys.}, B480:37--54, 1996.

\bibitem{Fujii:1998kn}
Yasunori Fujii et~al.
\newblock The nuclear interaction at oklo 2 billion years ago.
\newblock {\em Nucl. Phys.}, B573:377--401, 2000.

\bibitem{Lamoreaux:2003ii}
S.~K. Lamoreaux.
\newblock Neutron moderation in the oklo natural reactor and the time variation
  of alpha.
\newblock {\em Phys. Rev.}, D69:121701, 2004.

\bibitem{Petrov:2005pu}
Yu.~V. Petrov, A.~I. Nazarov, M.~S. Onegin, V.~Yu. Petrov, and E.~G.
  Sakhnovsky. Phys. Rev. C {\bf 74}, 064610 (2006); hep-ph/0506186


\bibitem{Lamur06}
Gould C.R., Sharapov E.I. and Lamoreaux S.K. Phys. Rev. C {\bf 74}, 024607 (2006)

\bibitem{ENDF}
National Nuclear Data Center homepage, http://www.nndc.bnl.gov/
\bibitem{Ruffenach}
Ruffenach J.C. Natural Fission Reactors. Proceedings of a meeting
of natural fission reactors. Paris, France (Decem. 1977), IAEA,
Vienna (1978). P.441.
\bibitem{MCNP}
J.F. Briesmeister Editor: "MCNP - A General Monte Carlo N-particle transport code".
LANL Report LA-13709-M (Apr. 2000)

\bibitem{MCU}
Majorov L.V., Gomin E.A., Gurevich M.I. Program complex of MCU-REA
with nuclear constant library DLC/MCUDAT-2.2. (unpublished)

\bibitem{Hidaka:1998}
H. Hidaka and P. Holliger, Geochimica et Cosmochimica Acta {\bf 62}, No.~1, 89 (1998)
\bibitem{Holliger}
P. Holliger, C. Devillers, G. Retali, in {\it Natural Fission
Reactors}, IAEA-TC-119 (IAEA, Vienna, 1978), IAEA-TC-119/20,
p.553.
\bibitem{Holliger2}
P. Holliger and C. Devillers, Earth Planet. Sci. Lett., {\bf 52},
76 (1981).
\bibitem{mure}
O. M$\acute e$plan, J. Wilson, A. Bidaud, S. David et al. MURE,
MCNP Utility for Reactor Evolution. User Guide - Version 1.0.
(2009). (unpublished)
\bibitem{Muhabhab}
S.F. Mughabghab. Atlas of Neutron Resonancs. Resonance Parameters
and Thermal CrossSection Z=1--100. $5^{th}$ edition. Elsevier,
2006.
\bibitem{Murphy}
 J.K. Webb, J.A. King, M.T. Murphy, V.V. Flambaum et al.
 ArXiv:1008.3907.
\bibitem{Levshakov}
S.A. Levshakov, P. Molaro, S. Lopez, S.D'Odorico, M. Centurion, P.
Bonifacio, I.I. Agafonova, and D. Reimers, astro-ph/0703042.
\bibitem{Ivanchik}
A.Ivanchik, P. Petitjean, D. Varshalovich, B. Araci et al. Astron.
Astrophys. {\bf 440}, 45 (2005); astro-ph/0507174.
\bibitem{Bekenstein}
J.D. Bekenstein, Phys. Rev. {\bf D25}, 1527 (1982)
\bibitem{Olive2002}
K.A. Olive, M. Pospelov, Phys. Rev. {\bf D65}, 085044 (2002); hep-ph/0110377
\bibitem{Will}
C.M. Will, Living Rev. Relativity {\bf 9}, 3 (2006); gr-qc/0510072
\bibitem{Dent}
T. Dent, JCAP 0701 (2007) 013; hep-ph/0608067

\bibitem{Copeland2004}
E.J. Copeland, N.J. Nunes, M. Pospelov, Phys. Rev. {\bf D69}, 023501 (2004)

\end{thebibliography}

\newpage

\end{document}